\documentclass[twocolumn]{aastex62}

\usepackage{epsfig}
\usepackage{epstopdf}
\usepackage{graphics}
\usepackage{xcolor}

\usepackage{array}
\usepackage{booktabs}
\usepackage{amsmath,amssymb,amsfonts,amsbsy}
\usepackage{mathrsfs}
\usepackage{url}
\usepackage{multirow}

\usepackage{graphicx}      
\usepackage{multirow}

\usepackage{color}
\usepackage{ulem}
\usepackage{enumerate}
\usepackage{arydshln}
\usepackage{graphicx}
\usepackage{array}
\usepackage{multirow}

\renewcommand{\vec}[1]{\boldsymbol{#1}}

\renewcommand{\figurename}{Fig.}
\renewcommand{\tablename}{Tab.}
\begin{document}

\title{Two kinds of Galactic source populations could explain the cosmic-ray observation up to the ``knee” region}

\author{Furong Li}
\affiliation{School of Space and Earth Sciences, Beihang University, Beijing 100191, China}

\author{Wei Liu}
\affiliation{State Key Laboratory of Particle Astrophysics, Institute of High Energy Physics, Chinese Academy of Sciences, Beijing 100049, China}
\email{liuwei@ihep.ac.cn}  

\author{Yali Shao}
\affiliation{School of Space and Earth Sciences, Beihang University, Beijing 100191, China}
\email{yshao@buaa.edu.cn}

\author{Yi-Qing Guo}
\affiliation{State Key Laboratory of Particle Astrophysics, Institute of High Energy Physics, Chinese Academy of Sciences, Beijing 100049, China}
\affiliation{University of Chinese Academy of Sciences, Beijing 100049, China}
\email{guoyq@ihep.ac.cn}

\begin{abstract}

Observations of diffuse gamma rays above hundreds of TeV from the Galactic disk provide strong evidence for the existence of PeV cosmic-ray accelerators—so-called PeVatrons—in the Galaxy. However, mounting observations have ruled out most supernova remnants as likely PeVatron candidates, suggesting instead that multiple populations of cosmic-ray sources exist in the Galaxy. Recently, the LHAASO collaboration reported the detection of ultra-high-energy gamma rays from microquasars, establishing that the black holes in these systems, which accrete matter from companion stars, are powerful PeV particle accelerators. In this work, we propose a two-component source model to explain the observed cosmic-ray spectra and composition up to the PeV range. Below approximately $100$ TeV, supernova remnants serve as the dominant sources; above this energy, microquasars are considered the primary candidate population. Within this scenario, the assumption of a charge-dependent cutoff well accounts for the latest measurements, including the proton and helium spectra up to the PeV range, the energy-dependent composition, and the all-particle spectrum. In contrast, the nuclei-dependent cutoff hypothesis is ruled out by the data.
\end{abstract}


\section{Introduction}
\label{sec:intro}
According to the theory of diffusive shock acceleration, cosmic rays (CRs) accelerated in the strong shock fronts follow a power-law rigidity spectrum, $\mathcal{R}^{-\nu}$, with an index of $\nu = 2$ in the strong shock limit. Following their escape from the sources, CRs execute random walk through the interstellar medium subsequently, a process commonly modeled as diffusion. The diffusion coefficient is typically assumed to depend solely on rigidity, i.e., $D(\mathcal{R}) \propto \mathcal{R}^{\delta}$, where $\delta \approx 0.3$–$0.6$. Consequently, the propagated CR spectrum is expected to follow a single power-law of the form $\phi \propto \mathcal{R}^{-\nu-\delta}$. Supernova remnants (SNRs) were widely accepted as the leading candidates for the Galactic cosmic rays responsible for accelerating cosmic rays to PeV energies.

However, this conventional picture has been challenged by extensive observations in recent years. First, the expected single power-law spectrum is not supported by the data. Nearly all measured energy spectra of primary and secondary cosmic-ray nuclei exhibit an excess above a rigidity of $\sim 200$ GV \citep{2011Sci...332...69A, 2017ApJ...839....5Y, 2021PhR...894....1A}. For both proton and helium, this excess extends up to tens of TeV, followed by a spectral softening \citep{2017JCAP...07..020A, 2017ApJ...839....5Y, 2019SciA....5.3793A, 2021PhRvL.126t1102A}. Moreover, secondary-to-primary ratios, such as boron-to-carbon and boron-to-oxygen, also show a hardening above tens of GeV. The origin of these spectral features can be broadly categorized into—though not limited to—three scenarios: acceleration mechanisms \citep{2011ApJ...729L..13O, 2012PhRvL.108h1104M, 2010ApJ...725..184B, 2011PhRvD..84d3002Y, 2017ApJ...835..229K}, propagation effects \citep{2012PhRvL.109f1101B, 
2012ApJ...752L..13T, 2012JCAP...01..010B, 2012ApJ...752...68V, 2014A&A...567A..33T, 2014ApJ...782...36E, 2015A&A...583A..95A, 2015PhRvD..92h1301T,  2016PhRvD..94l3007F, 2016ApJ...819...54G, 2016ChPhC..40a5101J, 2018ChPhC..42g5103G, 2018PhRvD..97f3008G},  and the presence of nearby sources \citep{2012A&A...544A..92B, 2012MNRAS.421.1209T, 2013MNRAS.435.2532T, 2013A&A...555A..48B, 2015RAA....15...15L, 2015ApJ...803L..15T, 2015ApJ...815L...1T, 2017PhRvD..96b3006L}.

Meanwhile, the long-standing assumption that SNRs serve as the dominant sources of Galactic CRs has been increasingly questioned. Observations of sub-PeV diffuse gamma-ray emission from the Galactic plane, along with gamma rays from point sources, provide compelling evidence for the existence of PeVatrons in the Galaxy. Yet, most known SNRs appear to accelerate cosmic rays only below several tens of TeV, suggesting the presence of extra Galactic CR accelerators. The LHAASO collaboration recently reported the detection of a few gamma-ray sources with maximal energy $E > 100$ TeV, five of which are associated with microquasars \citep{2021Natur.594...33C, 2025NSRev..12af496L}. Notably, the extended ultrahigh-energy emission from the central region of SS 433 coincides spatially with a giant gas cloud, indicating a hadronic origin. These findings demonstrate that microquasars can act as highly efficient particle accelerators, capable of reaching energies beyond the PeV scale and thereby contributing significantly to the CR flux around the ``knee" region.

The CR ``knee" characterizes a softening of the all-particle spectrum at $\sim $ PeV, where the spectral index changes from $-2.7$ to $-3.1$. Its origin has remained a subject of debate since its discovery nearly $70$ years ago \citep{1958NCim....8S.742H}. Proposed explanations for this spectral break include source-related mechanisms, such as an acceleration limit or confinement at the source \citep{2009A&A...497...17V, 2012MNRAS.427...91O, 2011MNRAS.414.1432T, 2018ApJ...854...57F}, contributions from nearby pulsars \citep{2010ApJ...710..958K}, and propagation effects \citep{2010NJPh...12c3044S, 2010ApJ...710..236S, 2018PhRvD..97l3005J}. A key distinction among these models is whether the spectral cutoff depends on the charge number ($Z$) or the nucleon number ($A$) of individual CR species—a prediction testable through composition measurements. Recently, the LHAASO WFCTA collaboration released new measurements of the proton and helium energy spectra around the knee region \citep{2025arXiv250514447T, 2025arXiv251105013L}, revealing knee-like structures at approximately $3$ PeV for protons and $7.5$ PeV for helium.

In this work, We propose a two-component model of Galactic cosmic-ray sources to explain the observations up to PeV energies. In this hybrid model, supernova remnants (SNRs) dominate the flux below $\sim 100$ TeV, while microquasars prevail at higher energies. A charge-dependent ($Z$-dependent) high-energy cutoff successfully accounts for the individual spectra of major components, the energy-dependent composition, and the all-particle spectrum across the entire energy range. In contrast, a mass-dependent ($A$-dependent) cutoff is inconsistent with the observed individual spectra and can be ruled out. The paper is organized as follows: Section 2 outlines the model, Section 3 presents the results, and Section 4 provides a discussion and conclusions.

\section{Model Description}
\subsection{CR Propagation}
It has been recognized in recent years that the propagation of CRs in the Milky Way should depend on the spatial locations, as inferred by the HAWC and LHAASO observations of extended $\gamma$-ray halos around pulsars \citep{2017Sci...358..911A} and the spatial variations of the CR intensities and spectral indices from Fermi-LAT observations \citep{2016PhRvD..93l3007Y, 2016ApJS..223...26A}. The spatially-dependent propagation (SDP) model was proposed to explain the observed hardening of CR spectra above $200$ GV \citep{2012ApJ...752L..13T, 2015PhRvD..92h1301T, 2016PhRvD..94l3007F, 2016ApJ...819...54G, 2018ApJ...869..176L, 2018PhRvD..97f3008G, 2020ChPhC..44h5102T}, diffuse gamma-ray distribution and the large-scale anisotropies \citep{2019JCAP...10..010L, 2019JCAP...12..007Q, 2021FrPhy..1624501Y}.

In the SDP model, the diffusive halo is divided into two parts: the inner and outer halo. Within the inner halo, the diffusion coefficient is much smaller than that in the outer halo, as indicated by the HAWC observations. The shape of the magnetic halo is usually approximated as a cylinder, with the Galactic disk embedded in the middle. The radial boundary of the magnetic halo is set to the Galactic radius, i.e., $R = 20~\mathrm{kpc}$, whereas its half-thickness $z_h$ needs to be evaluated by fitting the CR data \citep{PhysRevD.95.083007}. Both CR sources and the interstellar medium (ISM) are mainly distributed
within the Galactic disk, whose width $z_s$ is set to be $\sim 200~\mathrm{pc}$. Aside from the diffusion effect, CR nuclei may also undergo convection, reacceleration, and fragmentation due to collisions with interstellar gas. At lower energies, CR nuclei further lose energy via ionization and Coulomb scattering. The transport equation is generally written as
\begin{equation}
\begin{aligned}
\frac{\partial \psi(\mathbf{r}, p, t)}{\partial t}
=\, & Q(\mathbf{r}, p, t)
+ \nabla \cdot \left( D_{xx} \nabla \psi
- \mathbf{V}_c \psi \right) \\
& + \frac{\partial}{\partial p}
\left[
p^{2} D_{pp}
\frac{\partial}{\partial p}
\left(
\frac{1}{p^{2}} \psi
\right)
\right] \\
& - \frac{\partial}{\partial p}
\left[
\dot{p}\, \psi
- \frac{p}{3}
(\nabla \cdot \mathbf{V}_c)\, \psi
\right] \\
& - \frac{\psi}{\tau_f}
- \frac{\psi}{\tau_r} ,
\end{aligned}
\label{eq:CR_transport}
\end{equation}

where $\psi = dn/dp$ is the CR density per unit particle momentum $p$
at position $\mathbf{r}$,
$Q(\mathbf{r}, p, t)$ is the source function,
$D_{xx}$ and $D_{pp}$ are the diffusion coefficients in the spatial and
momentum spaces, respectively,
$\mathbf{V}_c$ is the convection velocity,
$\dot{p}$ is the momentum loss rate,
and $\tau_f$ and $\tau_r$ denote the fragmentation and radioactive decay
timescales.
At the boundary of the halo, free escape of CRs is assumed, namely,
$\psi(R, z, p) = \psi(r, \pm z_h, p) = 0$.

The spatial diffusion coefficient $D_{xx}$ can be parameterized as
\begin{equation}
D_{xx}(r,z, {\cal R} )= D_{0}F(r,z)\beta^{\eta} \left(\dfrac{\cal R}
{{\cal R}_{0}} \right)^{\delta_{0}F(r,z)},
\label{eq:diffusion}
\end{equation}
\begin{equation}
F(r,z) = \left\{
\begin{array}{ll}
{\dfrac{N_m}{1+f(r,z)}+\left[1-\dfrac{N_{m}}{1+f(r,z)}\right]}\left(\dfrac{{z}}{\xi{z}_{\rm h}} \right)^{n},  &  {{|z|} \leq \xi{z}_{\rm h}} \\
\\
1,  &  { {|z|} > \xi{z}_{\rm h}} \\
\end{array},
\right.
\end{equation}
where $r$ and $z$ are cylindrical coordinates, ${\cal R}$ is the particle's rigidity, $\beta$ is the particle's velocity in units of light speed, $D_0$ and $\eta$ are constants. The half-thickness of the propagation halo is $z_h$, and the
half-thickness of the inner halo is $\xi z_{h}$.

In this work, we adopt the diffusion reacceleration model, with the diffusive-reacceleration coefficient $D_{pp}$, which correlated with $D_{xx}$ via $D_{pp}D_{xx} = \dfrac{4p^{2}v_{A}^{2}}{3\delta(4-\delta^{2})
(4-\delta)}$, where $v_A$ is the Alfv\'en velocity, $p$ is the momentum,
and $\delta$ is the rigidity dependence slope of the diffusion coefficient \citep{1994ApJ...431..705S}. The numerical package DRAGON is used to solve the propagation equation of CRs \citep{2017JCAP...02..015E}.

\subsection{Background SNRs}
As discussed in Section 1, the background source should include two categories as SNRs and PeVatrons. For simplicity, the spatial distribution of them is approximated as an axisymmetric form, which can be parameterized as

\begin{equation}
f(r, z) = \left(\dfrac{r}{r_\odot} \right)^\alpha \exp \left[-\dfrac{\beta(r-r_\odot)}{r_\odot} \right] \exp \left(-\dfrac{|z|}{z_s} \right) ~,
\label{eq:radial_dis}
\end{equation}
where $r_\odot \equiv 8.5$ kpc represents the distance from the Galactic center to the solar system. Parameters $\alpha$ and $\beta$ are taken as $1.69$ and $3.33$ in this work \citep{1996A&AS..120C.437C}. The density of the SNR distribution decreases exponentially along the vertical height from the Galactic plane, with $z_{s} = 200$ pc. The injection spectrum of nuclei is assumed to be a broken power-law function of particle rigidity with an exponential cut off:
\begin{equation}
q({\cal R}) \propto \left(\dfrac{\cal R}{ {\cal R}_{\rm 0} }\right)^{-\nu} \exp \left( -\dfrac{\cal R}{{\cal R}_{\rm c}} \right) ~.
\end{equation}

\subsection{Local SNR}
The fine structure of spectral hardening and break-off at $\sim 200$ GeV and $\sim 14$ TeV, respectively, seems to be from a local source. The local source is also necessary to explain the evolution of anisotropy with energy. The influence of local sources on the large-scale dipole anisotropy has been studied in past studies, for example \citep{2006APh....25..183E, 2012JCAP...01..010B, 2013ApJ...766....4P, 2014ApJ...785..129K, 2013APh....50...33S, 2016PhRvL.117o1103A}. In this work, the progenitor of Geminga, an SNR, was introduced as the local source. The injection process of the SNR is approximated as a burst-like process. The source injection rate as a function of time and rigidity is assumed to be
\begin{equation}
Q({\cal R},t)=Q_{0}(t) \left(\dfrac{\cal R}{{\cal R}_0}
\right)^{-\gamma} \exp \left( -\dfrac{\cal R}{{\cal R}_{\rm c}}
\right) ~,
\label{eq:nearby}
\end{equation}
\begin{equation}
Q_{0}(t) = q_{0} \delta(t-t_0) ~, \\
\end{equation}
where  ${\cal R}_{\rm c}$ is the cutoff rigidity and $t_0$ is the time of the supernova explosion. The propagated spectrum from Geminga SNR is thus a convolution of the Green's function and the time-dependent injection rate $Q_0(t)$
\citep{1995PhRvD..52.3265A}

\begin{equation}
\varphi(\vec{r}, {\cal R}, t) = \int_{t_i}^{t} G(\vec{r}-\vec{r}^\prime, t-t^\prime, {\cal R}) Q_0(t^\prime) d t^\prime .
\end{equation}
{The normalization is determined through fitting Galactic cosmic rays energy spectra, which results in a total energy of $\sim 2.2\times 10^{50}$ erg for protons and $\sim 7.5\times 10^{49}$ erg for helium. If 10\% of kinetic energy is used to accelerate CRs, the total energy of the supernova explosion is estimated to $\sim 3\times 10^{51}$ erg.}

\subsection{Microquasars}
 Black holes (BH), one of the most intriguing objects in the universe, can manifest themselves through electromagnetic radiation initiated by the accretion flow. Some stellar-mass BHs drive relativistic jets when accreting matter from their companion stars, called microquasars. The accretion process of BHs may lead to various vibrant astrophysical phenomena over a wide range of radiation bands through converting the gravitational energy of falling matter or the rotational energy of BHs into energetic particles. Accreting BHs have long been suggested as efficient accelerators of particles exceeding 100 TeV\citep{1991ApJ...381..220A, 1998NewAR..42..579A, 2005A&A...432..609B}, and recent LHAASO observations of gamma-ray emission extending beyond 100~TeV from sources associated with several microquasars such as SS~433, GRS~1915+105, and V4641~Sgr strongly support the idea that they can act as Galactic PeVatrons \citep{2025arXiv251001369K}.

The spatial distribution of microquasars is assumed to follow that of massive stellar populations in the Galaxy, which is well represented by the distributions of high-mass X-ray binaries and SNRs\citep{2025arXiv251001369K, 2025arXiv250620193Z}. This treatment corresponds to a Galactic population of microquasars contributing to the high-energy component, in contrast to scenarios invoking a single dominant source such as Sgr~A$^\ast$\citep{2016Natur.531..476H}.
For simplicity, the source density is parameterized in an axisymmetric form similar to that adopted for SNRs:
\begin{equation}
f_{\rm MQ}(r, z) =\left(\dfrac{r}{r_\odot} \right)^\alpha \exp \left[-\dfrac{\beta(r-r_\odot)}{r_\odot} \right] \exp \left(-\dfrac{|z|}{z_s} \right) ~.
\label{eq:mq_spatial}
\end{equation}

The injection spectrum of cosmic rays accelerated by microquasars is assumed to have the same form as background SNRs. Since the exponential cutoff could be either charge number dependent or nucleon number dependent, we assume that the exponential cutoff is $\exp \left(-\dfrac{\cal R}{ {\cal R}_c} \right)$ and $\exp \left(-\dfrac{E}{E_c} \right)$, separately, where $\cal R$ and $E$ are rigidity and total energy.

Observational and modeling studies of microquasar jets indicate that the kinetic powers of these systems span a broad range, typically 
$L_{\rm jet}\sim10^{37}$--$10^{39}~{\rm erg~s^{-1}}$
\citep{1999A&A...351..156P,2007MNRAS.376.1341R,10.1093/pasj/psaf110}. 
For characteristic active lifetimes of $\sim10^{4}$--$10^{5}$ yr
\citep{particles7030047,10.1111/j.1365-2966.2011.18388.x} 
and a hadronic acceleration efficiency of a few percent
\citep{2003A&A...410L...1R,2014ApJ...783...91C}, 
the total energy injected into cosmic rays by a single microquasar can reach 
$\sim10^{47}$--$10^{50}$ erg. 
Recent studies based on LHAASO observations suggest that, under a hadronic interpretation, SS~433 may sustain a proton luminosity of order 
$\sim10^{38}~{\rm erg~s^{-1}}$
\citep{2024arXiv241008988T}. 
The cumulative proton injection luminosity from Galactic microquasars could plausibly be larger by about one order of magnitude. 
These estimates indicate that a population of order ten active microquasars can provide a total CR injection power of 
$\sim10^{38}$--$10^{39}~{\rm erg~s^{-1}}$
\citep{2025PhRvD.112l3015Z}. 
This level is well below the canonical Galactic CR luminosity of 
$\sim10^{41}~{\rm erg~s^{-1}}$.

\subsection{Bayesian Parameter Inference}
Bayesian inference has been widely used in astrophysics, cosmology, and particle physics for parameter estimation and uncertainty quantification\citep{2008ConPh..49...71T,2009A&A...497..991P,2011ApJ...729..106T}. In this work, we adopt a Bayesian approach to constrain the free parameters of our model using the available cosmic-ray data.

Given a parameter set $\boldsymbol{\theta}$ and the observational data set $D$, the posterior probability distribution is given by
\begin{equation}
P(\boldsymbol{\theta}\mid D) \propto \mathcal{L}(D \mid \boldsymbol{\theta}) \, \Pi(\boldsymbol{\theta}),
\end{equation}
where $\mathcal{L}(D\mid \boldsymbol{\theta}) $ is the likelihood function and $\Pi(\boldsymbol{\theta})$ is the prior distribution.
In this work, the parameters associated with the background SNRs component and the local SNR component are first determined in the stage-1 fit. In the stage-2 analysis, these parameters are fixed to their best-fit values, while the remaining high-energy parameters associated with the microquasars component are sampled with MCMC.

The likelihood function is defined as
\begin{equation}
\ln \mathcal{L} = -\frac{1}{2}\chi^2(\boldsymbol{\theta}),
\end{equation}
with
\begin{equation}
\chi^2(\boldsymbol{\theta})=
\sum_i
\left[
\frac{F_i^{\rm model}(\boldsymbol{\theta})-F_i^{\rm data}}
{\sigma_i}
\right]^2.
\end{equation}
where $F_i^{\rm data}$ and $\sigma_i$ are the observed flux and its corresponding uncertainty in the $i$th energy bin, respectively, while $F_i^{\rm model}(\boldsymbol{\theta})$ is the model prediction in the same bin.

We sample the posterior distribution using the affine-invariant MCMC sampler \texttt{emcee} \citep{2013PASP..125..306F}. Uniform priors are adopted for all free parameters within physically motivated ranges.

\section{Results}
In the SDP model, the propagation parameters include $D_0$, $N_m$, $\delta_0$, $\xi$, $\eta$, $v_A$, and $z_h$, which are constrained by fitting the boron-to-carbon (B/C) ratio, whose values are listed in Table \ref{tab:sdp}. The injection parameters of each nuclear species for the background SNRs include normalization $A_0$, power index $\nu$, and the cutoff rigidity ${\cal R}_{c, \rm SNR}$. For the local SNR, the corresponding parameters are injection power $q_0$, power index $\gamma$ and the cutoff rigidity ${\cal R}_{c, \rm local}$. The injection parameters for microquasars include normalization $A_0$, power index $\nu$, and the cutoff rigidity ${\cal R}_{c, \rm MQ}$ (or the cutoff energy $E_c$.  All of the above injection parameters are determined by fitting the individual energy spectrum.

\begin{figure}[!tb]
\centering
\includegraphics[width=0.45\textwidth]{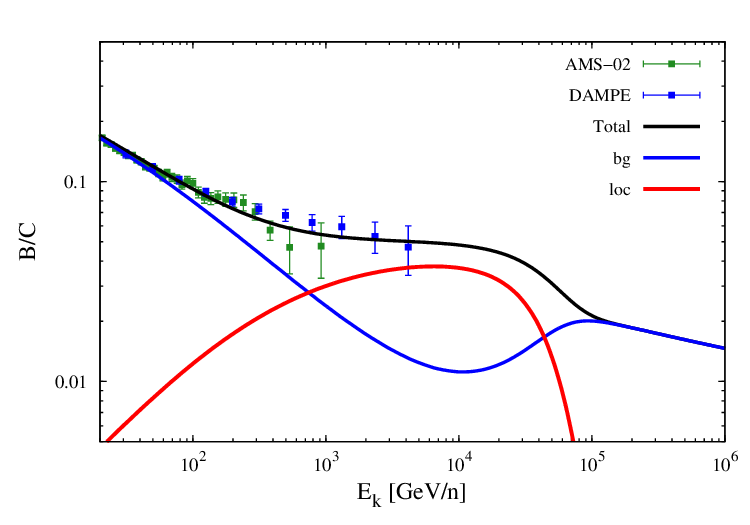}
\caption{Comparison between model calculations and observations for the ratios of boron to carbon. The data points are taken from AMS-02\citep{2021PhR...894....1A},
and DAMPE
\citep{2022SciBu..67.2162D}.
}
\label{fig:bcratio}
\end{figure}

In our model, secondary boron nuclei associated with the background SNRs and microquasars are produced by the spallation of primary cosmic-ray nuclei, mainly C, N, and O, during their diffusive propagation in the interstellar medium. In contrast, the boron from the local SNR is generated through interactions with the gas during the acceleration process within the source itself. The comparison between the modeled and measured B/C ratio is presented in Fig. \ref{fig:bcratio}. Contributions from background SNRs and microquasars, the local SNR, and their total are shown by the blue, red, and black lines, respectively. At energies below $10$ TeV, background boron nuclei are primarily sourced from background SNRs. As shown, the DAMPE measurement of the B/C ratio indicates a spectral hardening above $100$ GeV/n. This excess is explained by boron nuclei produced via the fragmentation of heavier nuclei during CR acceleration in the local SNR. Based on the observed softening of proton and helium spectra at tens of TeV, the acceleration limit of the local SNR is inferred to be at a rigidity of $\sim 35$ TV (Tab. \ref{tab:inj}). Therefore, the B/C ratio component from the local SNR is predicted to decrease above $\sim 10$ TeV. At energies beyond tens of TeV, boron production is dominated by primary cosmic rays from microquasars. Owing to their harder injection spectrum compared to SNRs, microquasar-generated boron becomes the dominant component over the background above $10$ TeV. The associated hardening of the B/C ratio is expected to persist up to $\sim 30$ TeV, where a softening appears. In the range from $100$ TeV to $1$ PeV, the B/C ratio is predicted to exhibit a power-law form. All these features await testing by future experiments.

\begin{figure}[!tb]
\centering
\includegraphics[width=0.45\textwidth]{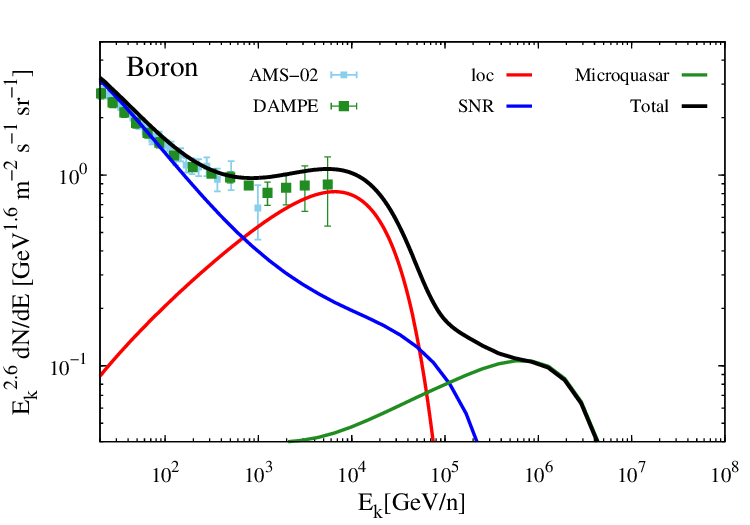}
\caption{The boron cosmic-ray spectra.The data are taken from
\citep{2018PhRvL.120b1101A} and DAMPE
\citep{2022SciBu..67.2162D}.
}
\label{fig:B}
\end{figure}

Figure~\ref{fig:B} shows the total boron spectrum together with the individual contributions from the background SNRs, the local SNR, and microquasars. The excess above $\sim 200$~GeV/n is mainly attributed to the local SNR component, which gradually decreases at energies of several tens of TeV/n. At higher energies, above $\sim 100$~TeV/n, the microquasar component becomes increasingly important.

\begin{table*}[htbp]
\centering
\caption{Transport Parameters}
\setlength{\tabcolsep}{13pt} 
\renewcommand{\arraystretch}{1.2} 
\normalsize                        
\begin{tabular}{@{}ccccccc@{}}
\toprule
$D_0\,(10^{28}\,\mathrm{cm^2\,s^{-1}})$ &
$N_m$ &
$\delta_{\mathrm{0}}$ &
$\xi$ &
$\eta$ &
$v_A\,(\mathrm{km\,s^{-1}})$ &
$z_h\,({\rm kpc})$
\\
\midrule
7.9  & 0.45 & 0.56 & 0.1 & 0.05 & 6 & 5\\
\bottomrule
\end{tabular}
\label{tab:sdp}
\end{table*}

\newcommand{\rownums}[9]{#1 & #2 & #3 & #4 & #5 & #6 & #7 & #8 & #9 \\}
\newcommand{\NA}{---}  

\begin{table*}[htbp]
\centering
\caption{Injection parameters of major elements for background SNRs, local SNR, and microquasars.}
\label{tab:inj}
\setlength{\tabcolsep}{2.5pt}
\small
\begin{tabular}{l c c c c c c c c c}
\toprule
& \multicolumn{3}{c}{\textbf{comp1}} & \multicolumn{3}{c}{\textbf{local}} & \multicolumn{3}{c}{\textbf{comp2}} \\
\cmidrule(lr){2-4}\cmidrule(lr){5-7}\cmidrule(lr){8-10}
\textbf{Element} &
$A_0\,[\mathrm{m^{-2}\,sr^{-1}\,s^{-1}\,GeV^{-1}}]$ & $\nu$ & ${\cal R}_{c, \rm SNR}\,[\mathrm{TV}]$ &
$q_0\,[\mathrm{GeV^{-1}}]$ & $\gamma$ & ${\cal R}_{c, \rm local}\,[\mathrm{TV}]$ &
$A_0\,[\mathrm{m^{-2}\,sr^{-1}\,s^{-1}\,GeV^{-1}}]$ & ${\nu}_{\rm MQ}$ & ${\cal R}_{c,MQ}\,[\mathrm{PV}]$ \\
\midrule
P  & \rownums{$4.15\times10^{-2}$}{2.45}{350}{$4.3\times10^{52}$}{2.15}{35}{$1.40\times10^{-3}$}{2.05}{5}
He & \rownums{$2.30\times10^{-3}$}{2.31}{350}{$5.6\times10^{51}$}{2.02}{35}{$1.06\times10^{-4}$}{2.01}{5}
C  & \rownums{$9.22\times10^{-5}$}{2.42}{350}{$2.0\times10^{51}$}{2.15}{35}{$4.00\times10^{-6}$}{2.05}{5}
N  & \rownums{$1.39\times10^{-5}$}{2.42}{350}{$2.2\times10^{50}$}{2.15}{35}{$6.50\times10^{-7}$}{2.05}{5}
O  & \rownums{$1.07\times10^{-4}$}{2.42}{350}{$1.2\times10^{51}$}{2.15}{35}{$5.25\times10^{-6}$}{2.05}{5}
Ne & \rownums{$1.91\times10^{-5}$}{2.42}{350}{$3.6\times10^{50}$}{2.15}{35}{$7.88\times10^{-7}$}{2.05}{5}
Mg & \rownums{$2.41\times10^{-5}$}{2.42}{350}{$1.3\times10^{50}$}{2.15}{35}{$1.01\times10^{-6}$}{2.05}{5}
Si & \rownums{$2.33\times10^{-5}$}{2.42}{350}{$1.2\times10^{50}$}{2.15}{35}{$1.06\times10^{-6}$}{2.05}{5}
Fe & \rownums{$2.26\times10^{-5}$}{2.42}{350}{$2.0\times10^{50}$}{2.15}{35}{$7.94\times10^{-7}$}{2.05}{5}
\bottomrule
\end{tabular}
\vspace{0.4em}
\begin{minipage}{0.96\linewidth}
\footnotesize
\emph{Notes.} $A_0$ is the differential flux normalization; $R_c$ is the cutoff rigidity.
\end{minipage}
\end{table*}

\begin{figure}[!tb]
    \centering
    \begin{minipage}{0.45\textwidth}
        \centering
        \includegraphics[width=\textwidth]{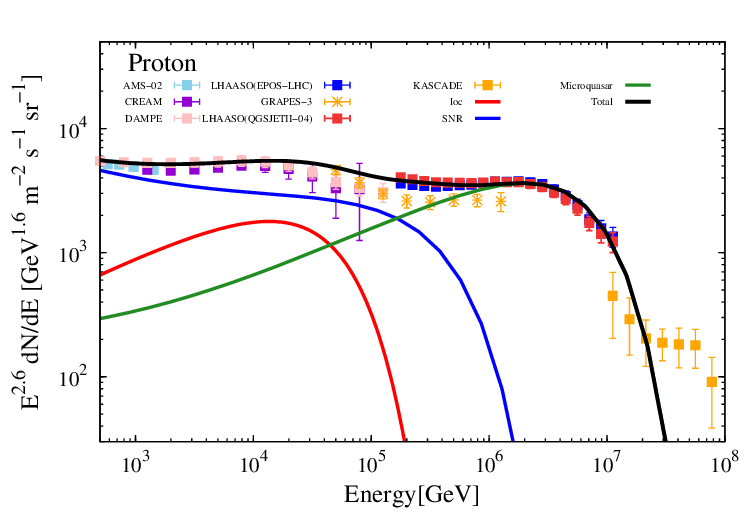}
    \end{minipage}
    \hfill
    \begin{minipage}{0.45\textwidth}
        \centering
        \includegraphics[width=\textwidth]{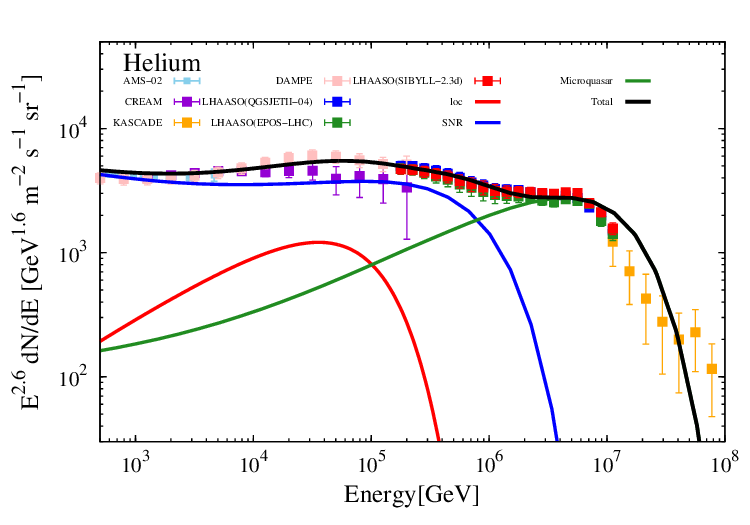}
    \end{minipage}
    \caption{The proton (upper) and helium (lower) cosmic-ray spectra. The data are taken from LHAASO\citep{2025arXiv250514447T, 2025arXiv251105013L}, AMS-02\citep{2021PhR...894....1A}, CREAM\citep{2017ApJ...839....5Y}, DAMPE\citep{2019SciA....5.3793A, 2025arXiv251105409D}, GRAPES-3\citep{2024PhRvL.132e1002V}and KASCADE\citep{2013APh....47...54A} .}
    \label{fig:prot_heliumz}
\end{figure}

Fig. \ref{fig:prot_heliumz} illustrates the proton and helium energy spectra taking into account three source components. The blue, red, and green lines are the contributions from background SNRs, local SNR, and microquasars, respectively. The black line is the sum of them. The hardening of the proton and helium nuclei spectra above $200$ GV is principally due to the local SNR contribution, which then softens around $\sim$ tens of TeV. The cutoff is consistent with the assumption of rigidity dependence. The whole SNRs in the Galaxy contribute to the cosmic rays up to the rigidity $\sim 350$ TV. This is because the available gamma-ray observations found that the cutoff of gamma-ray spectra from most of the SNRs is at tens of TeV. Above hundreds of TeV, the major contribution of CRs originates from the microquasars. But at lower energies, the contribution of the microquasars is less. 
In this work, we adopt a harder injection spectrum for microquasars, motivated by previous theoretical studies and recent observational results\citep{2026ApJ...997..163Y,2025PhRvD.112l3015Z}. Specifically, we assume moderately hard injection indices of $\sim2.05$ for protons and $\sim2.01$ for helium. This reflects the expectation that particle acceleration in microquasar jet environments may not follow the same mechanisms as non-relativistic shocks in SNRs. Previous theoretical studies have shown that processes such as magnetic reconnection and shear acceleration in relativistic jet systems can produce hard spectra close to $\nu \simeq 2$ or even harder\citep{2007Ap&SS.309..119R,2014PhRvL.113o5005G,2004ApJ...617..155R,2022ApJ...933..149R}. Recent LHAASO observations and modeling of microquasar systems, such as SS~433, suggest efficient acceleration of particles up to the PeV range and favor a hard particle spectrum with a spectral index $\nu \sim 2$, supporting the use of relatively hard injection spectra for the microquasar component \citep{10.1093/nsr/nwaf496}. Here, to explain the drop of protons and heliums at PeV energies, the cutoff is assumed to be rigidity dependent and the cutoff rigidity is $5$ PV.

\begin{figure}[!tb]
\centering
\includegraphics[width=0.50
\textwidth]{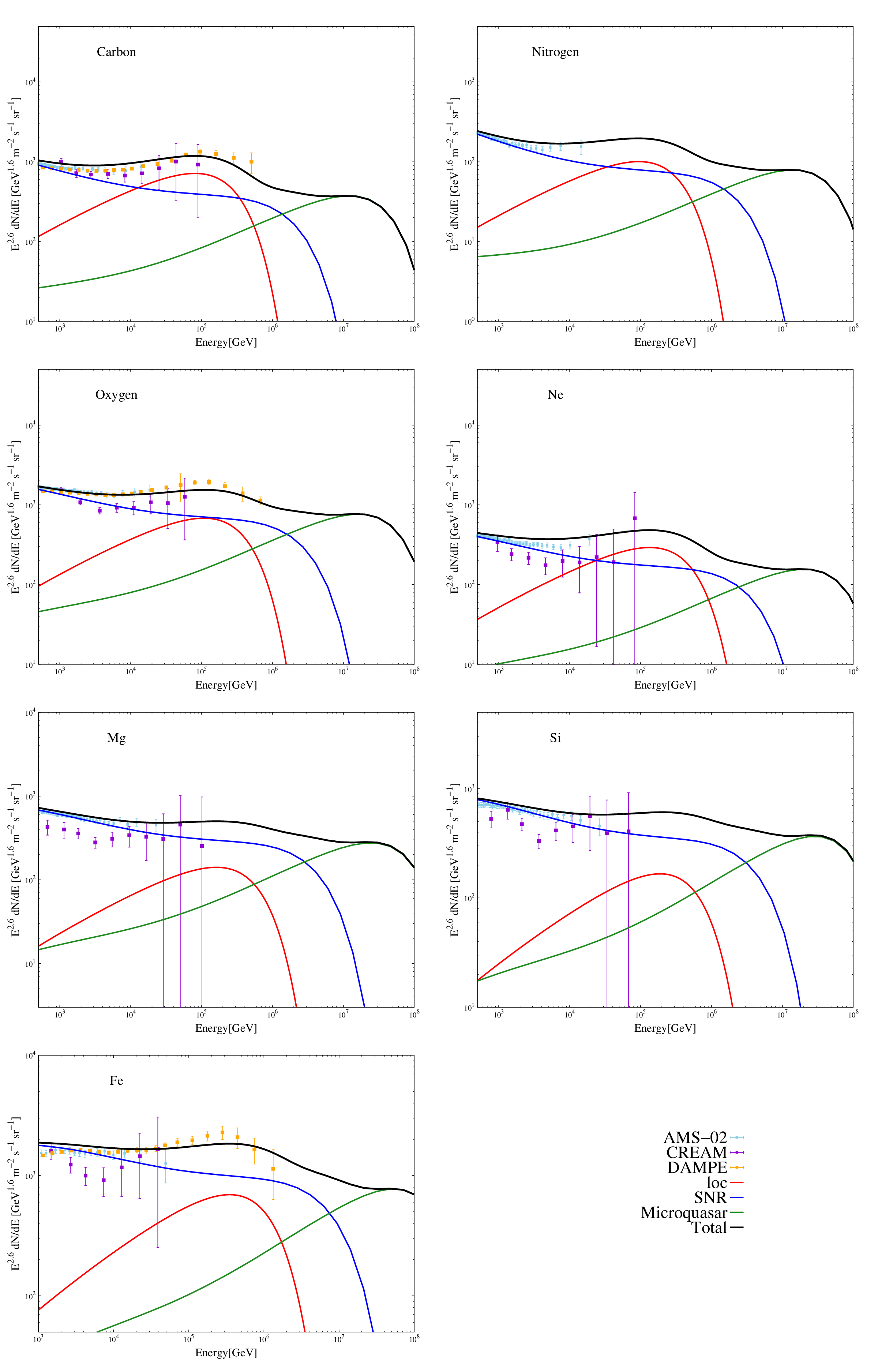}
\caption{The individual species of C, O, Ne, Mg, Si, and Fe are shown in panels.The data are taken from AMS-02\citep{2021PhR...894....1A,2021PhRvL.126d1104A}  CREAM\citep{2009ApJ...707..593A} and DAMPE\citep{2025arXiv251105409D} .}
\label{fig:CNO}
\end{figure}

\begin{figure}[!tb]
    \centering
    \begin{minipage}{0.5\textwidth}
        \centering
        \includegraphics[width=\textwidth]{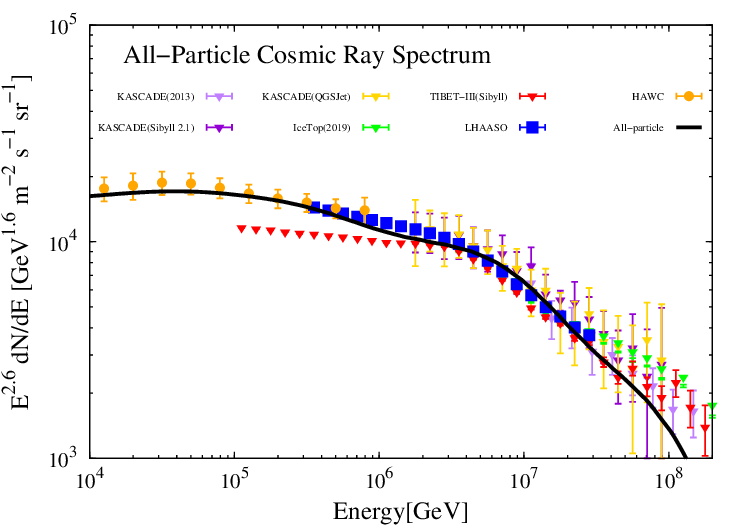}
    \end{minipage}
    \hfill
    \begin{minipage}{0.5\textwidth}
        \centering
        \includegraphics[width=\textwidth]{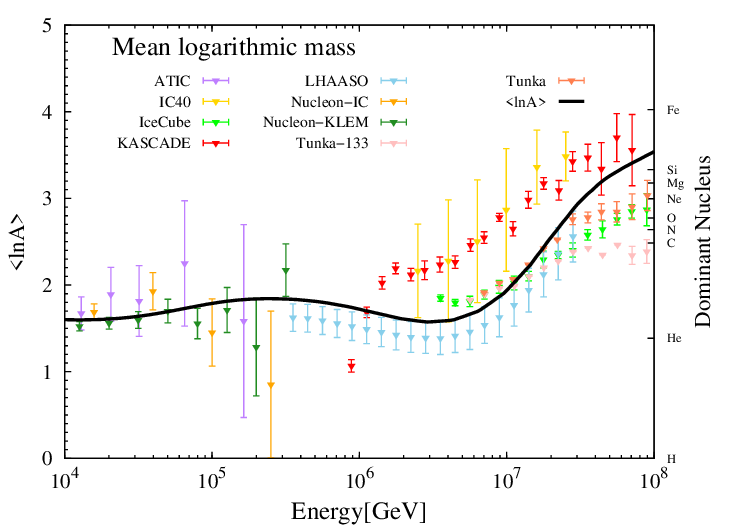}
    \end{minipage}
    \caption{Upper pane: The all-particle cosmic-ray spectrum, the data are taken from LHAASO\citep{2024PhRvL.132m1002C}, KASCADE\citep{2024JCAP...05..125K}, KASCADE(sibyll2.1)\citep{2005APh....24....1A}, KASCADE(QGSJet)\citep{2005APh....24....1A}, IceTop(2019)\cite{2019PhRvD.100h2002A}, TIBET\citep{2008ApJ...678.1165A} and HAWC\citep{2017PhRvD..96l2001A}.
    Lower panel: The distribution of $\langle \ln A \rangle$, the data are taken from LHAASO\citep{2024PhRvL.132m1002C}, ATIC\citep{2009BRASP..73..564P}, IC40\citep{2013APh....42...15I}, IceCube\citep{2019PhRvD.100h2002A}, KASCADE\citep{2024JCAP...05..125K}, Nucleon-IC\citep{2019AdSpR..64.2546G}, Nucleon-KLEM\citep{2019AdSpR..64.2546G}, Tunka-133\citep{2012NIMPA.692...98B}, and Tunka\citep{2013APh....50...18B}.}
    \label{fig:allparticle_lnAz}
\end{figure}

The expected spectra of the heavier nuclei are shown in Fig. \ref{fig:CNO}. Since the measurements of the heavier nuclei above $100$ TeV are absent, the CR fluxes from each source component are mainly constrained by the all-particle spectrum and the energy dependence of composition, which are shown in Fig. \ref{fig:allparticle_lnAz}, as well as the measurements at lower energy. The all-particle spetrum and mean logarithmic mass is consistent with the latest measurements of LHAASO and HAWC. The injected parameters of heavier nuclei are listed in Tab. \ref{tab:inj}. The spectral indices of heavier nuclei are softer than helium.

\begin{figure*}[!tb]
    \centering
    \begin{minipage}{0.45\textwidth}
        \centering
        \includegraphics[width=\textwidth]{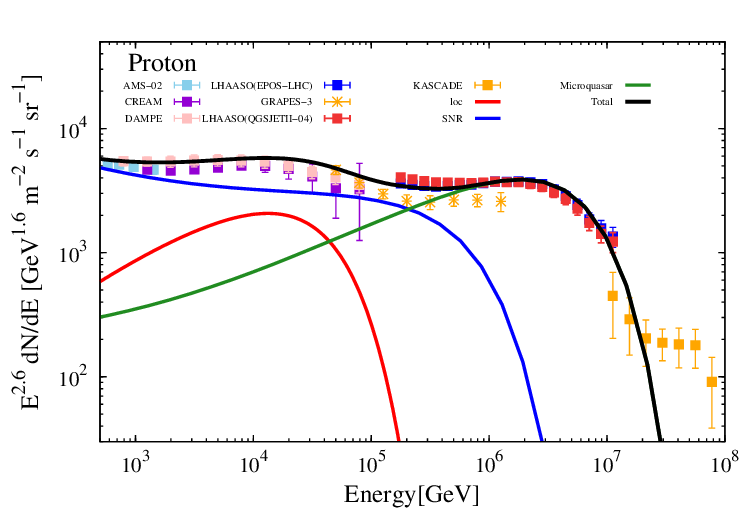}
    \end{minipage}
    \hfill
    \begin{minipage}{0.45\textwidth}
        \centering
        \includegraphics[width=\textwidth]{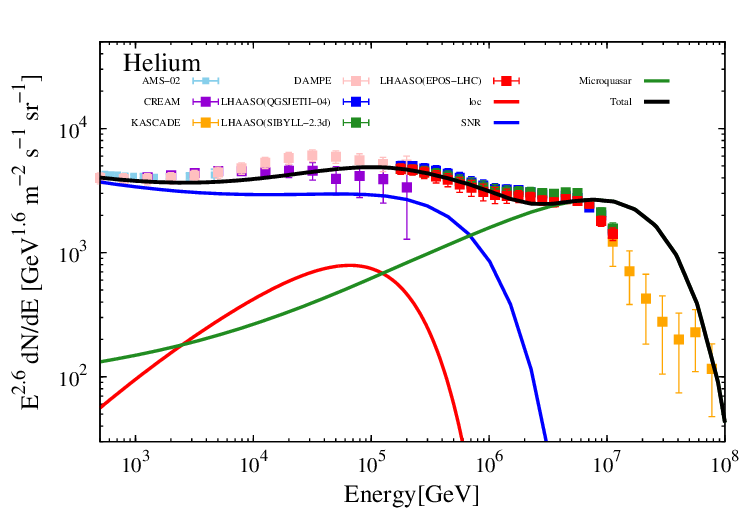}
    \end{minipage}
    \caption{The proton and helium cosmic-ray spectra.Left pane: Proton. Right panel: Helium. The data are taken from LHAASO\citep{2025arXiv250514447T, 2025arXiv251105013L}, AMS-02\citep{2021PhR...894....1A}, CREAM\citep{2017ApJ...839....5Y}, DAMPE\citep{2019SciA....5.3793A, 2025arXiv251105409D}, GRAPES-3\citep{2024PhRvL.132e1002V}and KASCADE\citep{2013APh....47...54A} .}
    \label{fig:prot_heliuma}
\end{figure*}

\begin{figure*}[!tb]
    \centering
    \begin{minipage}{0.45\textwidth}
        \centering
        \includegraphics[width=\textwidth]{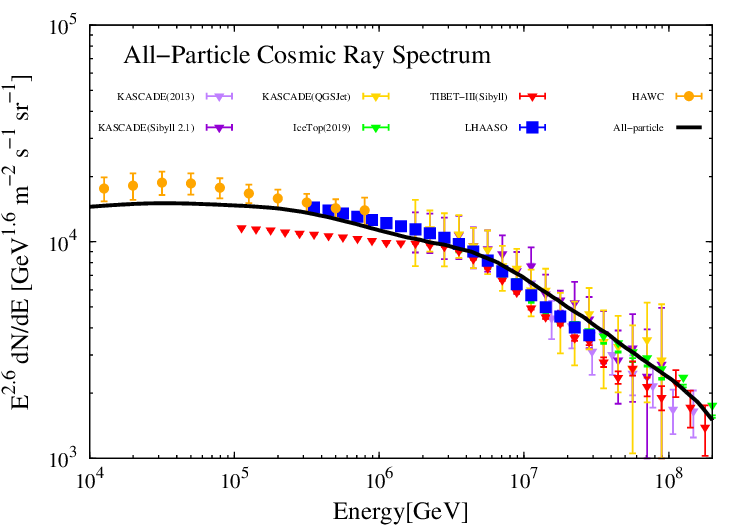}
    \end{minipage}
    \hfill
    \begin{minipage}{0.45\textwidth}
        \centering
        \includegraphics[width=\textwidth]{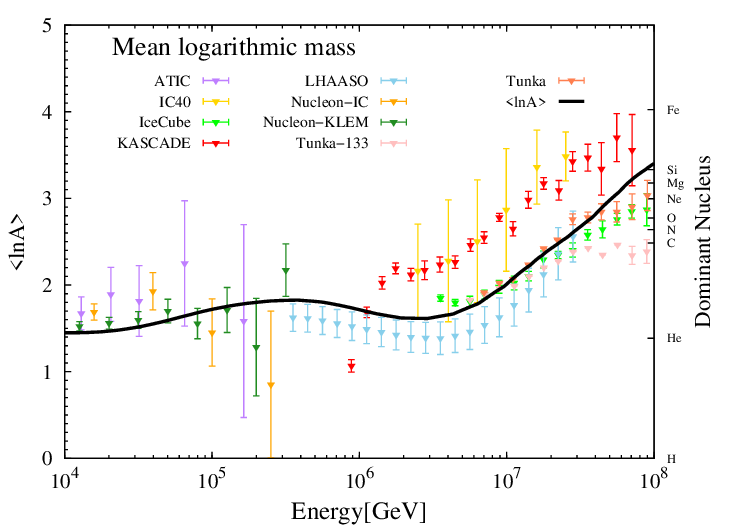}
    \end{minipage}
    \caption{Left pane: The all-particle cosmic-ray spectrum, the data are taken from LHAASO\citep{2024PhRvL.132m1002C}, KASCADE\citep{2024JCAP...05..125K}, KASCADE(sibyll2.1)\citep{2005APh....24....1A}, KASCADE(QGSJet)\citep{2005APh....24....1A}, IceTop(2019)\cite{2019PhRvD.100h2002A}, TIBET\citep{2008ApJ...678.1165A} and HAWC\citep{2017PhRvD..96l2001A} .
    Right panel:The distribution of $\langle \ln A \rangle$, the data are taken from LHAASO\citep{2024PhRvL.132m1002C}, ATIC\citep{2009BRASP..73..564P}, IC40\citep{2013APh....42...15I}, IceCube\citep{2019PhRvD.100h2002A}, KASCADE\citep{2024JCAP...05..125K}, Nucleon-IC\citep{2019AdSpR..64.2546G}, Nucleon-KLEM\citep{2019AdSpR..64.2546G}, Tunka-133\citep{2012NIMPA.692...98B}, and Tunka\citep{2013APh....50...18B}.}
    \label{fig:allparticle_lnAa}
\end{figure*}

\begin{figure}[!tb]
\centering
\includegraphics[width=0.45\textwidth]{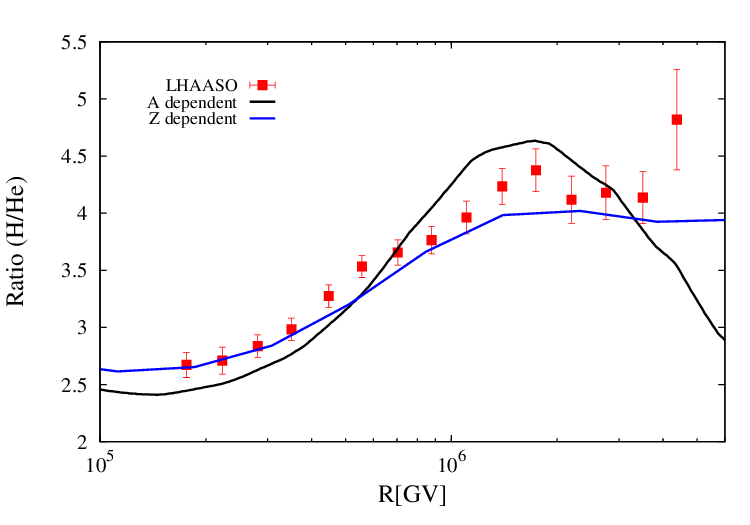}
\caption{The proton to helium flux ratio (H/He). The data are taken from LHAASO\citep{2025arXiv251105013L}.}
\label{fig:ratio}
\end{figure}

The proton and helium spectra with an $A$-dependent cutoff are shown in Fig. \ref{fig:prot_heliuma}. In this scenario, the knee-like structure of helium appears at a higher energy. Table \ref{fig:chi2} compares the fits of $Z$- and $A$-dependent cutoffs for the proton and helium knees, using spectra derived separately from the QGSJET-II-04 and EPOS-LHC hadronic models. In both cases, the $Z$-dependent description of the knee region performs significantly better than the $A$-dependent one. The corresponding all-particle spectrum and the mean logarithmic mass under the $A$-dependent assumption are shown in Fig. \ref{fig:allparticle_lnAa}. Notably, in the $A$-dependent case, the mean logarithmic mass varies more slowly above $30$ PeV (beyond the LHAASO measurement range) compared to the $Z$-dependent case. This is because the helium spectrum cutoff occurs at a higher energy under $A$-dependence, leading to a more gradual change in $\ln A$ with energy. As shown in Fig. \ref{fig:ratio}, the energy dependence of the proton-to-helium ratio differs between the $Z$- and $A$-dependent cutoff scenarios. Above $\sim 2$ PeV, the ratio exhibits a flat trend, and the $Z$-dependent model shows better consistency with the data.

\begin{table*}[!tb]
\centering
\caption{Chi-square per degree of freedom ($\chi^{2}/\mathrm{dof}$) comparison for proton and helium under $Z$-dependent and $A$-dependent scenarios.}
\begin{tabular*}{1.0\textwidth}{@{\extracolsep{\fill}}lcccc}
\toprule
Model & \multicolumn{2}{c}{Proton} & \multicolumn{2}{c}{Helium} \\
\cmidrule(lr){2-3} \cmidrule(lr){4-5}
       & $Z$-dependent & $A$-dependent & $Z$-dependent & $A$-dependent \\
\midrule
QGSJETII-04 & 0.87 & 2.32 & 1.39 & 5.02 \\
EPOS-LHC    & 0.44 & 0.59 & 2.04 & 4.17 \\
\bottomrule
\end{tabular*}
\label{fig:chi2}
\end{table*}

\begin{figure*}[htbp]
    \centering
    \includegraphics[width=0.95\textwidth]{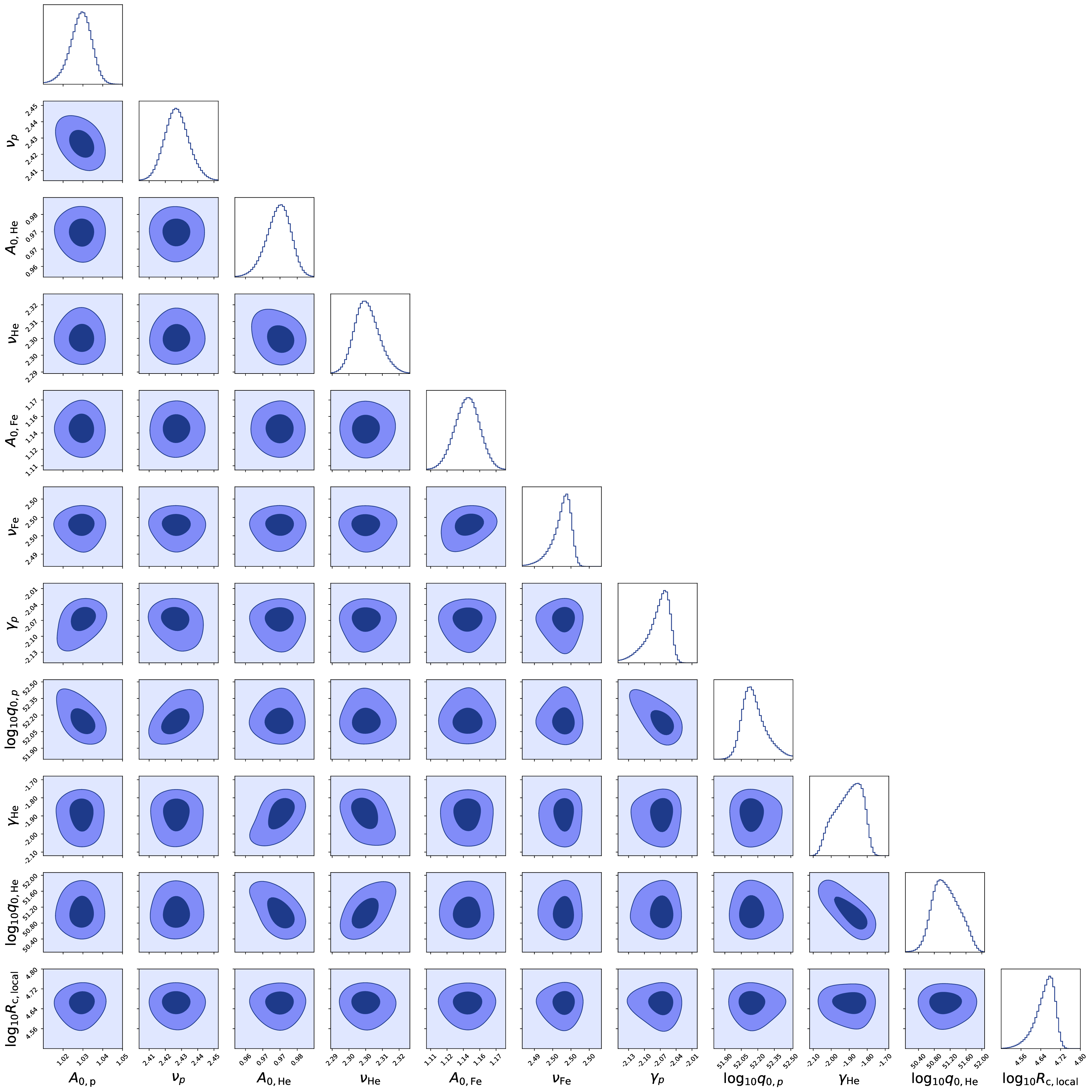}
    \caption{Posterior distributions and correlations of the background SNRs and local SNR parameters}
    \label{fig:mcmc}
\end{figure*}

\begin{figure*}[htbp]
    \centering
    \includegraphics[width=0.95\textwidth]{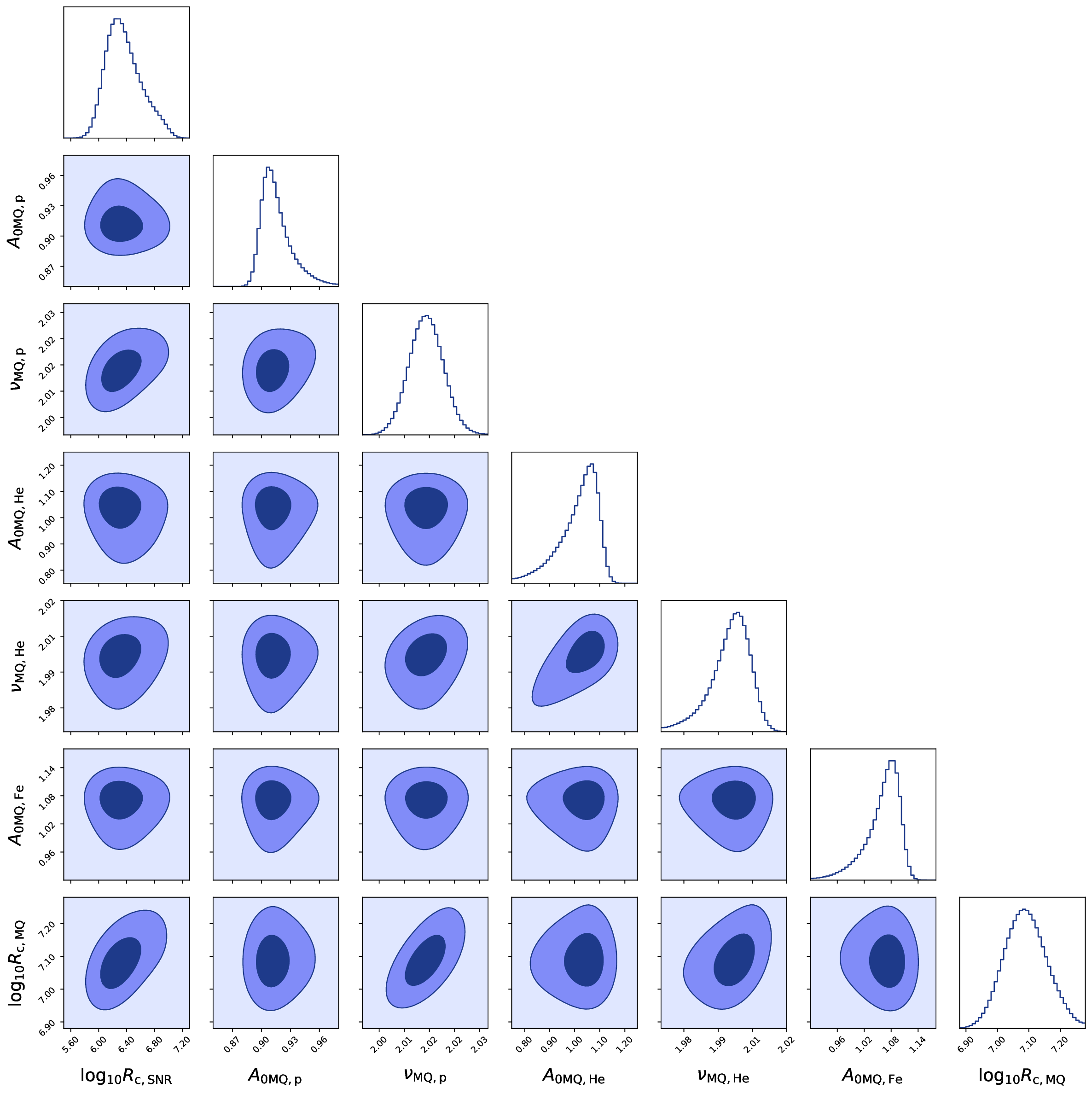}
    \caption{Posterior distributions of the microquasar parameters}
    \label{fig:stage2}
\end{figure*}

To investigate the parameter degeneracies and model robustness, we performed a two-stage MCMC analysis of the source injection parameters. In this analysis, the SDP propagation parameters are fixed to the values constrained mainly by the B/C ratio. The corresponding posterior distributions are shown in Figures~\ref{fig:mcmc} and \ref{fig:stage2}. The diagonal panels display the one-dimensional marginalized posterior distributions, while the off-diagonal panels show the correlations between parameters.

The source parameters are divided into two groups according to the energy ranges in which the corresponding components dominate. The local SNR component mainly contributes to the CR spectra below several tens of TeV, whereas the microquasar component becomes important in the all-particle spectrum above $\sim 100$~TeV. Therefore, in the stage-1 fit, we constrain the background SNRs and local SNR parameters using the composition and all-particle spectra at lower energies. In the stage-2 fit, the low-energy source parameters are fixed to their best-fit values, and the cutoff rigidity of the background SNRs together with the microquasar parameters are constrained using the all-particle spectrum above $\sim 100$~TeV.

Figure~\ref{fig:mcmc} presents the posterior distributions for the background SNRs and local SNR parameters obtained in the stage-1 fit. Mild negative correlations are observed between the normalization and spectral index, such as $A_0^{p}$--$\nu^{p}$ for the background SNR component and $q_0^{p}$--$\gamma^{p}$ for the local SNR component. Weak cross-correlations are also visible between the background SNRs and local SNR parameters (e.g., $A_0^{p}$–$q_0^{p}$ and $\nu^{p}$–$q_0^{p}$), indicating a limited level of compensation between the two source components in the energy range where both contribute.

Figure~\ref{fig:stage2} shows the posterior distributions of the high-energy parameters obtained in the stage-2 fit. Correlations between the microquasar normalization parameters and spectral indices are present but differ among species. For example, a weak positive correlation is visible for the helium component ($A_{0,MQ}^{\rm He}$--$\nu_{0,MQ}^{\rm He}$), while no clear correlation is found for the proton component. A clearer positive correlation is observed between the microquasar spectral index and cutoff rigidity (e.g., $\nu_{0,MQ}^{p}$--${\cal R}_{c, \rm MQ}$), indicating that both parameters affect the high-energy spectral shape. Correlations between different nuclear species and between different source components are generally weak.

Overall, although some mild-to-moderate correlations are present, no severe degeneracy is found, and the posterior distributions remain reasonably constrained.

The best-fit parameters inferred from the posterior distributions are summarized in Table~\ref{tab:bestfit}. For each parameter, we quote the median of the marginalized posterior distribution as the central value, and the 16th and 84th percentiles as the bounds of the $68\%$ credible interval.

\begin{table}[htbp]
\centering
\caption{Best-fit parameters and 68\% credible intervals derived from the posterior distributions}
\label{tab:bestfit}
\begin{tabular}{lll}
\midrule
Component & Parameter & 68\% limits \\
\midrule
\multicolumn{3}{c}{\textit{Stage-1: background SNRs and local SNR parameters}} \\
\midrule
\multirow{6}{*}{\shortstack[l]{comp1 \\ (background SNRs)}}
& $A_0^{p}$ & $1.0292^{+0.005}_{-0.004}$ \\
& $\nu^{p}$ & $2.4259^{+0.007}_{-0.007}$ \\
& $A_0^{\rm He}$ & $0.9720^{+0.003}_{-0.004}$ \\
& $\nu^{\rm He}$ & $2.3042^{+0.005}_{-0.006}$ \\
& $A_0^{\rm Fe}$ & $1.1438^{+0.010}_{-0.011}$ \\
& $\nu^{\rm Fe}$ & $2.4988^{+0.002}_{-0.001}$ \\
\cmidrule(lr){1-1}
\multirow{5}{*}{\shortstack[l]{comp2 \\    (Local SNR)}}
& $\gamma^{p}$ & $2.0694^{+0.014}_{-0.027}$ \\
& $\log_{10}(q_{0}^p)$ & $52.1520^{+0.128}_{-0.078}$ \\
& $\gamma^{\rm He}$ & $1.8963^{+0.066}_{-0.088}$ \\
& $\log_{10}(q_{0}^{\rm He})$ & $51.0787^{+0.375}_{-0.289}$ \\
& $\log_{10}(R_{\rm c,local})[{\rm GV}]$ & $4.6638^{+0.0268}_{-0.044}$ \\
\midrule
\multicolumn{3}{c}{\textit{Stage-2: microquasars parameters}} \\
\midrule
\multirow{1}{*}{\shortstack[l]{comp1}}
& $\log_{10}(R_{\rm c,SNR})[{\rm GV}]$ & $6.3266^{+0.280}_{-0.192}$ \\
\cmidrule(lr){1-1} 
\multirow{6}{*}{\shortstack[l]{comp3 \\ (Microquasars)}}
& $A_0^{p}$ & $0.9115^{+0.019}_{-0.009}$ \\
& $\nu^{p}$ & $2.0248^{+0.004}_{-0.004}$ \\
& $A_0^{{\rm He}}$ & $1.0333^{+0.049}_{-0.098}$ \\
& $\nu^{\rm He}$ & $1.9915^{+0.009}_{-0.006}$ \\
& $A_0^{\rm Fe}$ & $1.0709^{+0.022}_{-0.045}$ \\
& $\log_{10}(R_{\rm c,MQ})[{\rm GV}]$ & $7.0890^{+0.068}_{-0.063}$ \\
\midrule
\end{tabular}
\vspace{0.4em}
\begin{minipage}{0.96\linewidth}
\footnotesize
\emph{Notes.} unit of normalization $A_0^p/ A_0^{\rm He}/ A_0^{\rm Fe}$ and $q_{0}^p/q_{0}^{\rm He}$ are
$\mathrm{GeV^{-1}\, m^{-1}\, s^{-1}\, sr^{-1}}$ and 
$\mathrm{GeV^{-1}\, s^{-1}}$ respectively.
\end{minipage}
\end{table}

\section{Conclusion}
The LHAASO collaboration recently reported proton and helium energy spectra from $\sim 150$ TeV to $\sim 12$ PeV, measured with the WFCTA. Both spectra show a distinct hardening followed by a marked softening in the PeV range, revealing individual knee structures. Meanwhile, the traditional paradigm that SNRs could accelerate CRs to PeV energies has been challenged by the theoretical models and multi-wavelength data. Consequently, additional Galactic CR sources have to be invoked to account for the observed knee.

Recent detections of ultra-high-energy gamma-ray sources suggest that microquasars could be the potential Galactic PeVatrons. In this work, we propose a scenario where SNRs dominate the CR flux below $\sim 100$ TeV, while microquasars are introduced to account for the observed individual species spectra and the all-particle spectrum at the knee. We also examine the energy dependence of the mean logarithmic mass. Specifically, we explore the origin of the CR knee by investigating the energy dependence of the knee-like structures found in the proton and helium spectra.

In this model, the CR spectrum from microquasars is required to be harder than that from SNRs to successfully reproduce the knee regions in both the individual species and all-particle spectra, as well as the evolution of the mean logarithmic mass—a requirement consistent with gamma-ray observations. Meanwhile, the inferred injection spectrum of helium is harder than those of protons and heavier nuclei, similar to the trend observed at lower energies. Comparing the fits to the proton and helium knees under $Z$- and $A$-dependent scenarios, we find that the $Z$-dependent model provides a significantly better description of both the individual spectra and the flattening of the proton-to-helium ratio. This preference points to an acceleration or propagation origin for the cosmic-ray knee. While both $Z$- and $A$-dependent scenarios can explain the mean logarithmic mass within the LHAASO energy range, their predictions diverge at higher energies, with the $Z$-dependent model predicting a larger $\ln A$ at several PeV. Future measurements by experiments such as LHAASO at higher energies are needed to further clarify the origin of the knee.

\section*{Acknowledgements}
This work is supported by the National Natural Science Foundation of China (Nos. 12333006, 12375107, 12575120). Y.S. acknowledges support from the National Natural Science Foundation of China (NSFC) with grant No. 12303001, and the Fundamental Research Funds for the Central Universities.

\clearpage

\bibliographystyle{aasjournal}
\bibliography{refs}

\end{document}